\documentclass[12pt,a4paper]{article}

\usepackage[latin2]{inputenc}
\usepackage[T1]{fontenc}

\usepackage{pslatex}
\usepackage{ae,aecompl}
\usepackage{epsfig}
\usepackage{graphics}
\usepackage{graphicx}
\usepackage{amsmath}
\usepackage{amssymb}
\usepackage{pifont}
\usepackage{psfrag}

\usepackage{graphicx}
\usepackage{dcolumn}
\usepackage{bm}

 \def\ccc#1;#2{\left\langle #1
\left\vert #2 \right.\right\rangle} 

\begin{document}

\title{The Epps effect revisited}

\author{Bence T\'oth\footnote{E-mail: bence@maxwell.phy.bme.hu} $^{1,2}$ \and  J\'anos Kert\'esz$^{2,3}$}

\maketitle

$^1$ ISI Foundation - Viale S. Severo, 65 - I-10133 Torino, Italy 

$^2$ Department of Theoretical Physics, Budapest University of Technology and 
Economics - Budafoki \'ut. 8. H-1111 Budapest, Hungary

$^3$ Laboratory of Computational Engineering, Helsinki University of Technology 
-  P.O.Box 9203, FI-02015, Finland

\begin{abstract}
We analyse the dependence of stock return cross-correlations on the
sampling frequency of the data known as the Epps effect: For high
resolution data the cross-correlations are significantly smaller than
their asymptotic value as observed on daily data. The former
description implies that changing trading frequency should alter the
characteristic time of the phenomenon. This is not true for the
empirical data: The Epps curves do not scale with market activity. The
latter result indicates that the time scale of the phenomenon is
connected to the reaction time of market participants (this we denote
as human time scale), independent of market activity.  In this paper
we give a new description of the Epps effect through the decomposition
of cross-correlations. After testing our method on a model of
generated random walk price changes we justify our analytical results
by fitting the Epps curves of real world data.

\end{abstract}

\section{Introduction}
\label{intro}
1979 Epps reported results showing that stock return correlations
decrease as the sampling frequency of data increases
\cite{epps1979}. Since his discovery the phenomenon has been detected
in several studies of different stock markets
\cite{bonanno2001,zebedee2001,tumminello2006} and foreign exchange markets
\cite{lundin1999,muthuswamy2001}.

Cross-correlations between the individual assets are the main factors
in classical portfolio management thus it is important to understand
and give their accurate description on different time scales. This is
especially so, since today the time scale in adjusting portfolios to
events occuring may be in the order of minutes.


Considerable effort has been devoted to uncover the phenomenon found by Epps
\cite{reno2003,iori2004,iori2006,kwapien2004,zhang2006,bouchaud2005}. 
However most of the works aim to construct a better statistical
measure for co-movements in prices in order to exclude bias of the
estimator by microstructure effects
\cite{Barndorff-NielsenShephard2005,CorsiAudrino2007,HayashiYoshida2005,Martens2003,Andersen2006,AndersenBollerslev1998,JongNijman1997,BallTorous2000},
only a few searching for the description of the microstructure
dynamics.

Up to now two main factors causing the effect have been revealed: The first one is a
possible lead-lag effect between stock returns
\cite{lo1990_1,kullmann2002,toth2006} which can appear mainly between stocks
of very different capitalisation and/or for some functional
dependencies between them. In this case the maximum of the
time-dependent cross-correlation function can be found at non zero
time lag, resulting in increasing cross-correlations as the sampling
time scale gets into the same order of magnitude as the characteristic
lag.  This factor can be easily understood, morever, in a recent study
\cite{toth2006} we showed that through the years this effect becomes
less important as the characteristic time lag shrinks, signalising an
increasing efficiency of stock markets. As the Epps effect can also be
found for the case when no lead-lag effect is present, in the
following we will focus only on other possible factors.

The second, more important factor is the asynchronicity of ticks in case of
different stocks \cite{reno2003,iori2004,lo1990_1,lo1990_2}. Empirical results
\cite{reno2003} showed that taking into account only the synchronous ticks
reduces to a great degree the Epps effect, i.e.  measured correlations
on short sampling time scale increase. Naturally one would expect that
for a given sampling frequency growing activity decreases the
asynchronicity, leading to a weaker Epps effect. Indeed Monte Carlo
experiments showed an inverse relation between trading activity and
the correlation drop
\cite{reno2003}.

However, the analysis of empirical data showed \cite{toth2007} that
the explanation of the effect solely by asynchronicity is not
satisfactory.  After eliminating the effect of changing asymptotic
cross-correlations through the years (scaling with the asymptotic
value), the curves of cross-correlation as a function of sampling time
scale tend to collapse to one curve and surprisingly we do not find a
measurable reduction of the characteristic time of the Epps effect,
while the trading frequency grew by a factor of $\sim 5-10$ in the
period. These results will be discussed further in details in Section
\ref{activity}.

The characteristic time of market phenomena can usually be split up
into three kinds of market time scales: the frequency of trading on
the market (which we will denote as \textit{activity}), market
periodicities and the reaction time of traders to news, events. In
Ref. \cite{toth2007} we showed that the characteristic time of the
Epps effect does not scale with changing market activity (this we will
discuss in Section \ref{activity}), which points out that the
characteristic time of the Epps effect can not be determined solely by
the market activity causing asynchronicity. Market periodicities in
high frequency data are the different types of patterns, which can be
found in intraday data, as well as on broader time scales (see
e.g. Refs. \cite{Wood1985} and \cite{Chan1995}). Market periodicities
and intraday structure do not have a role in our results since we are
averaging them out. Hence we believe that the characteristic time of
the Epps effect is the outcome of a human time scale present on the
market: The time that market participants need to react to certain
pieces of news. There are several studies in the literature about
reaction time. The issue is connected both to behavioural finance
questions and to market efficiency. 1970, Fama defined an efficient
market as one in which prices fully reflect all available information
\cite{Fama1970}. This response to information in practice can not
happen instantaneously. There are several results reporting that
prices incorporate news within five to fifteen minutes after news
announcements
\cite{Dann1977,PatellWolfson1984,JenningsStarks1985,BarclayLitzenberger1988,Kim1997}. 
More recent studies showed similar results on the time that traders
needed to react to news \cite{Busse2002,Chordia2005,Chordia2008}.

Supposing that the Epps effect is possibly the outcome of a human time
scale present on the market motivated us to separate the terms in the
cross-correlation function, in order to study their behaviour one by
one. In this paper we suggest an analytic decomposition of the
cross-correlation function of asynchronous events using time lagged
correlations. As a second step we demonstrate the efficiency of the
formalism on the example of generated data.  Finally we describe and
fit the empirically observed dependence of the cross-correlations. We
find that the origin of the independence of the characteristic time of
the Epps effect on the trading frequency is the presence of a human
time scale in the time lagged autocorrelation
functions. Ref. \cite{bouchaud2005} already called the attention to
the importance of lagged cross-influences of stock returns in
explaining the Epps-effect.  Using a somewhat different formalism,
here we investigate thoroughly this relationship.

The paper is built up as follows: in Section \ref{empirical} we present the
data used and discuss the problems of the former descriptions. Section
\ref{theory} the decomposition of the cross-correlation coefficient, Section
\ref{model} shows a simulation model demonstrating the idea. In Section
\ref{application} we present the assumptions concerning real data and show 
fits and statistics for the Epps curves. We finish the paper with a
discussion.

\section{Empirical analysis}
\label{empirical}

\subsection{Data and methodology}
\label{data}
In our analysis we used the Trade and Quote (TAQ) Database of the New York
Stock Exchange (NYSE) for the period of 4.1.1993 to 31.12.2003, containing
tick-by-tick data. The data used was adjusted for dividends and splits.

We computed
the logarithmic returns of stock prices:
\begin{eqnarray}
\label{eq:ret}
r_{\Delta t}^{A}(t)=\ln \frac{p^{A}(t)}{p^{A}(t-\Delta t)},
\end{eqnarray}
where \(p^{A}(t)\) stands for the price of stock \textit{A} at time
\(t\). The prices were determined using previous tick estimator on the
high frequency data, i.e. prices are defined constant between two
consecutive trades.  The time dependent cross-correlation function
\(C_{\Delta t}^{A/B}(\tau)\) of stocks
\textit{A} and \textit{B} is defined by

\begin{eqnarray}
\label{eq:C}
C_{\Delta t}^{A/B}(\tau)=\frac{\left\langle r_{\Delta t}^{A}(t)r_{\Delta
t}^{B}(t+\tau)\right\rangle - \left\langle r_{\Delta t}^{A}(t)\right\rangle
\left\langle r_{\Delta
t}^{B}(t+\tau)\right\rangle}{\sigma^{A}\sigma^{B}}.\end{eqnarray}
The notion
\(\left\langle \cdots\right\rangle\) stands for the time average over the
considered period:
\begin{eqnarray}\label{eq:time_ave}
\left\langle r_{\Delta t}(t)\right\rangle =\frac{1}{T-\Delta t}\sum_{i=\Delta t}^{T} r_{\Delta t}(i),
\end{eqnarray}
where time is measured in seconds and \textit{T} is the time span of
the data.

\begin{figure}[htb!]
\begin{center}
\psfrag{Dt}[t][b][4][0]{$\Delta t$ [sec]}
\psfrag{rho_KO/PEP^Dt}[b][t][4][0]{$\rho_{\Delta t}^{KO/PEP}$}
\includegraphics[angle=-90,width=0.50\textwidth]{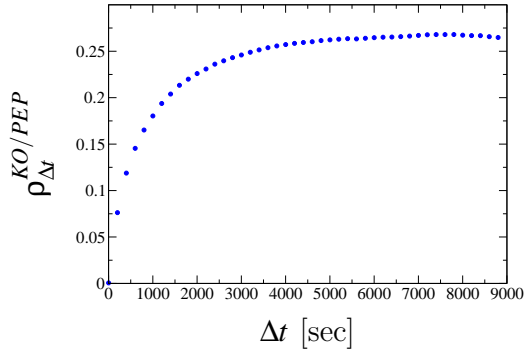}
\caption{The cross-correlation coefficient as a function of sampling time scale for
  the period 1993--2003 for the Coca-Cola Pepsi pair. Several hours
  are needed for the correlation to reach its asymptotic value.}
\label{fig:example}
\end{center}
\end{figure}

The standard deviation \(\sigma\) of the returns reads as:
\begin{eqnarray}\label{eq:sigma}\sigma=\sqrt{\left\langle r_{\Delta
t}(t)^{2}\right\rangle - \left\langle r_{\Delta
t}(t)\right\rangle^2},
\end{eqnarray}
both for $A$ and $B$ in \eqref{eq:C}. We computed correlations for
each day separately and averaged over the set of days, this way
avoiding large overnight returns and trades out of the market opening
hours. For pairs of stocks with a lead--lag effect the function
$C_{\Delta t}^{A/B}$ has a peak at non-zero $\tau$. The equal-time
cross-correlation coefficient is naturally: \(\rho_{\Delta
t}^{A/B}\equiv C_{\Delta t}^{A/B}(\tau=0)\). In our notations the Epps
effect means the decrease of \(\rho_{\Delta t}\) as \(\Delta t\)
decreases (see Figure
\ref{fig:example}). Since the prices are defined as being constant between two
consecutive trades, the \(\Delta t\) time scale of the sampling can be
chosen arbitrarily.

As stated above, we do not want to discuss the Epps effect originated
from the lead-lag effect in the correlations. Thus we consider only
pairs of stocks where the latter effect is negligible, i.e., for which
the price changes are highly correlated with the peak position of
$C_{\Delta t}^{A/B}$ of Equation
\eqref{eq:C} being at $\tau\approx 0$. 
The results shown in this paper can be generalised for all stock pairs
(though in case of an empirical study one can never fit all data). To
illustrate our results we will present results for some stock pairs
and in Section \ref{application} we will show statisitcs for a broader
set of data. The stocks mentioned in the paper are the following: Avon
Products, Inc. (AVP), Caterpillar Inc. (CAT), Colgate-Palmolive
Company (CL), E.I. du Pont de Nemours \& Company (DD), Deere \& Company
(DE), The Walt Disney Company (DIS), The Dow Chemical Company (DOW),
General Electric Co. (GE), International Business Machines
Corp. (IBM), Johnson \& Johnson (JNJ), The Coca-Cola Company (KO), 3M
Company (MMM), Motorola Inc. (MOT), Merck
\& Co., Inc. (MRK), PepsiCo, Inc. (PEP), Pfizer Inc. (PFE), 
The Procter \& Gamble Company (PG), Sprint Nextel Corp. (S), Vodafone
Group (VOD), Wal-Mart Stores Inc. (WMT).

\subsection{Time evolution of the characteristic time}
\label{activity}
Previous studies claimed the asynchronicity of ticks for different
stocks as the main cause of the Epps effect
\cite{reno2003,iori2004}. It is natural to assume that, for a given
sampling frequency, increasing trading activity should enhance
synchronicity, leading to a weaker Epps effect.

\begin{figure}[htb!]
\begin{center}
\psfrag{Dt}[t][b][4][0]{$\Delta t$ [sec]}
\psfrag{rho_CAT/DE^Dt}[b][t][4][0]{$\rho_{\Delta t}^{CAT/DE}$}
\psfrag{rho_KO/PEP^Dt}[b][t][4][0]{$\rho_{\Delta t}^{KO/PEP}$}
\psfrag{rho_MRK/JNJ^Dt}[b][t][4][0]{$\rho_{\Delta t}^{MRK/JNJ}$}
\includegraphics[angle=-90,width=0.50\textwidth]{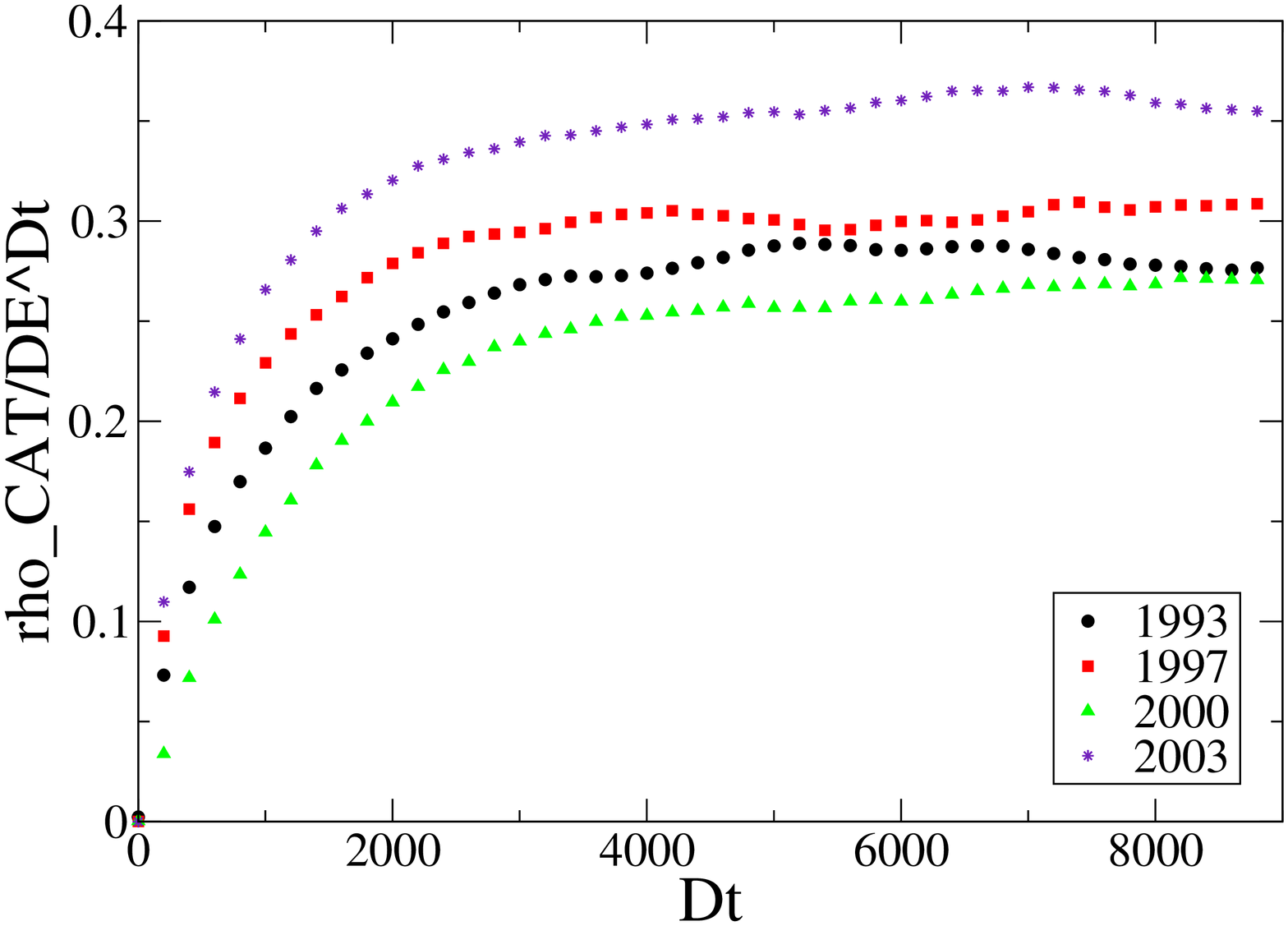}  
\includegraphics[angle=-90,width=0.50\textwidth]{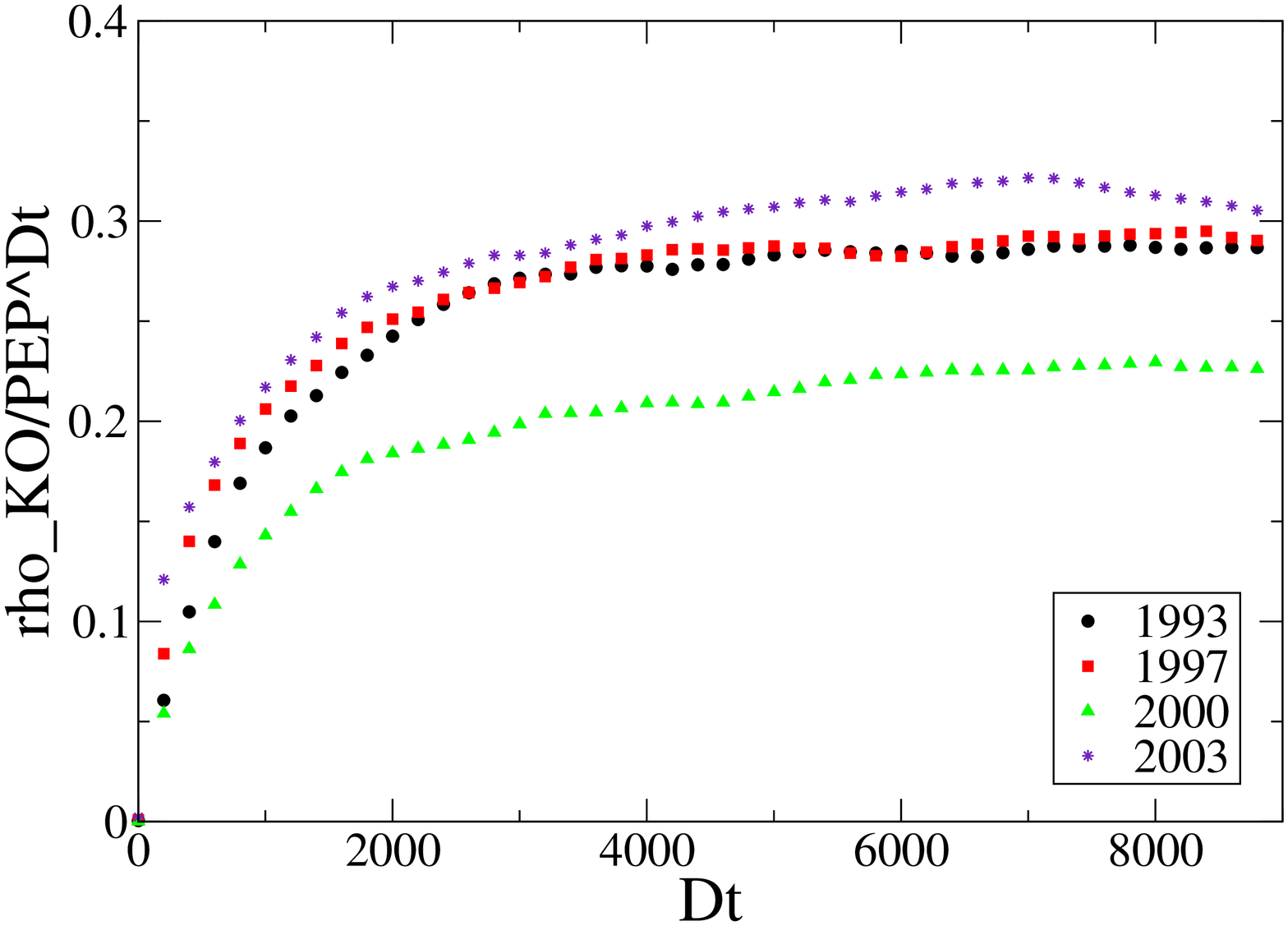} 
\includegraphics[angle=-90,width=0.50\textwidth]{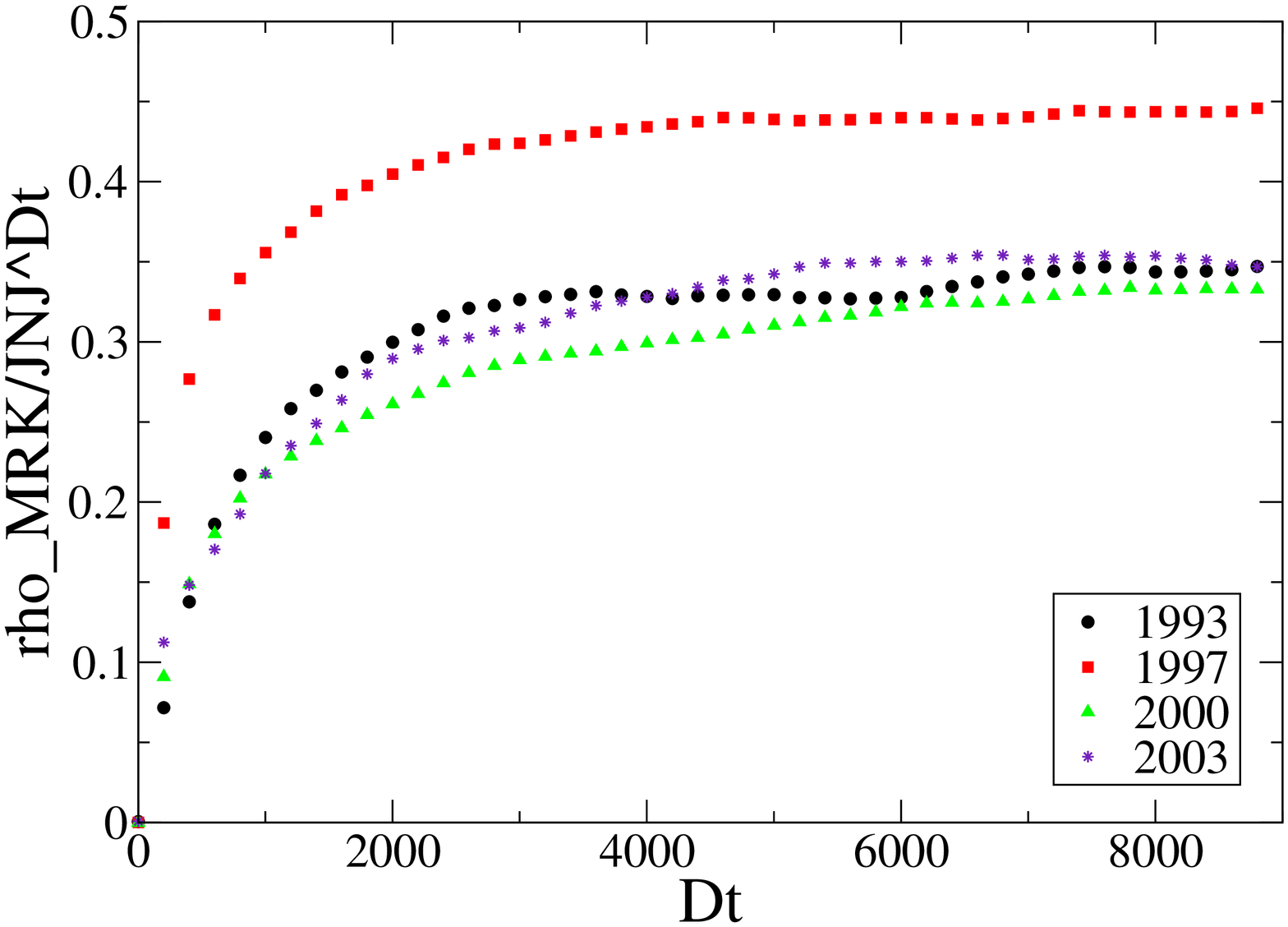}
\caption{The Epps curves for the CAT/DE (top), KO/PEP (middle) and
  MRK/JNJ (bottom) pairs for the years 1993, 1997, 2000 and 2003. The
  asymptotic value of the cross-correlations varies in time.}
\label{fig:epps_years}
\end{center}
\end{figure}

To study the trading frequency dependence of the cross-correlation
drop, we computed the Epps curve separately for different years. In
Figure \ref{fig:epps_years} the cross-correlation coefficients can be
seen as a function of the sampling time scale for the years 1993,
1997, 2000 and 2003 for three example stock pairs.

\begin{figure}[htb!]
\begin{center}
\psfrag{Dt}[t][b][4][0]{$\Delta t$ [sec]}
\psfrag{rho_CAT/DE^Dt}[b][t][4][0]{scaled $\rho_{\Delta t}^{CAT/DE}$}
\psfrag{rho_KO/PEP^Dt}[b][t][4][0]{scaled $\rho_{\Delta t}^{KO/PEP}$}
\psfrag{rho_MRK/JNJ^Dt}[b][t][4][0]{scaled $\rho_{\Delta t}^{MRK/JNJ}$}
\includegraphics[angle=-90,width=0.50\textwidth]{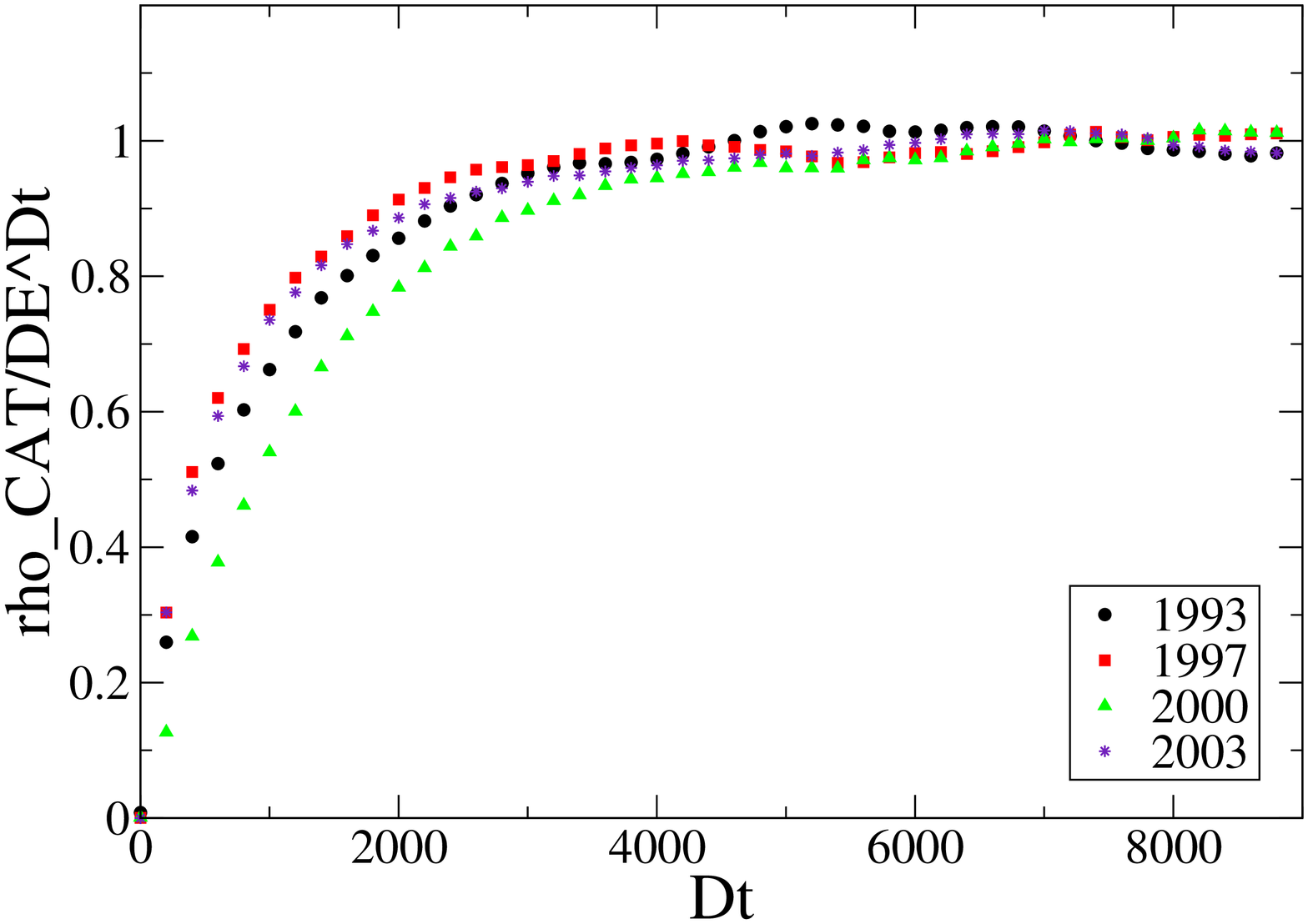}  
\includegraphics[angle=-90,width=0.50\textwidth]{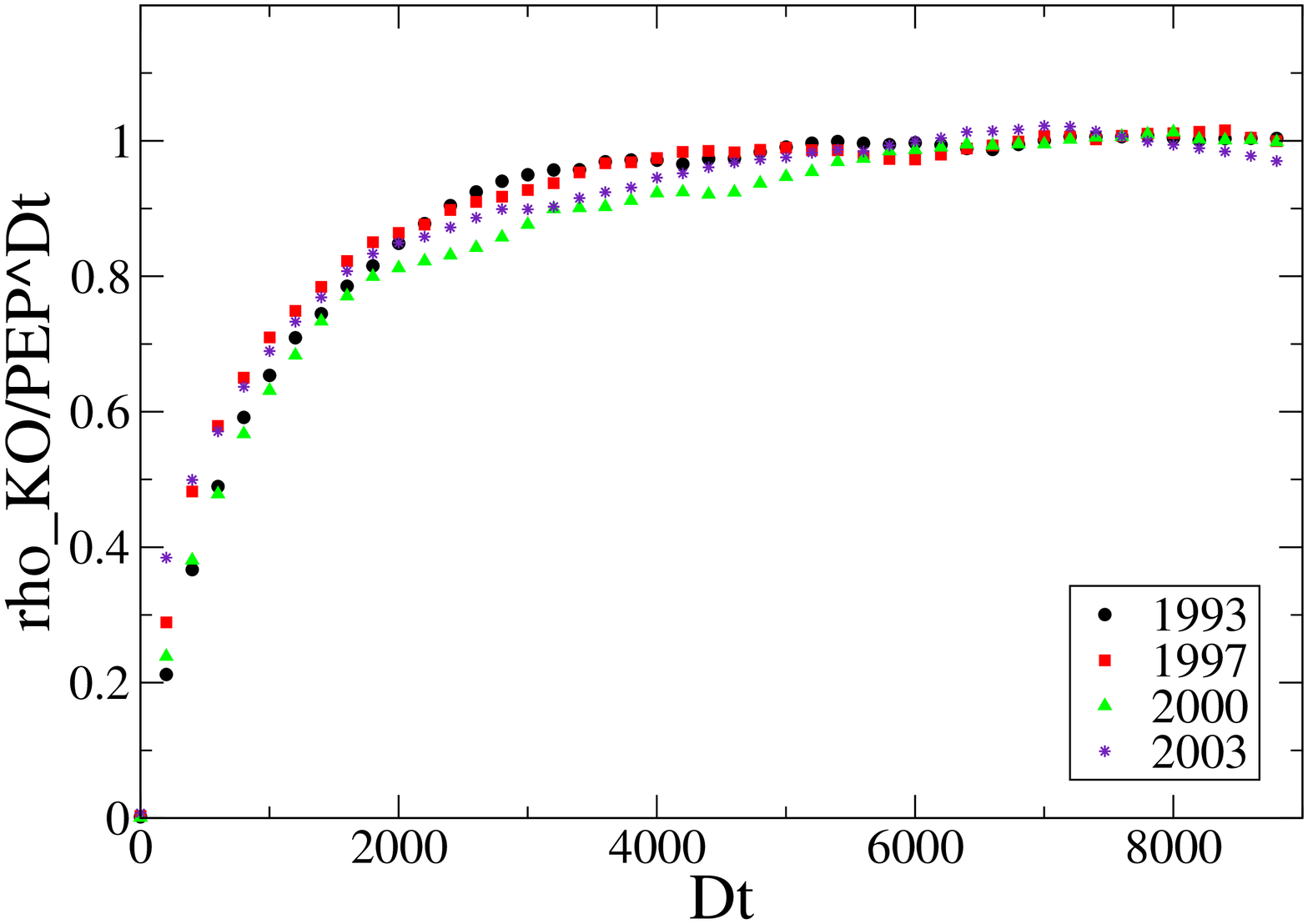} 
\includegraphics[angle=-90,width=0.50\textwidth]{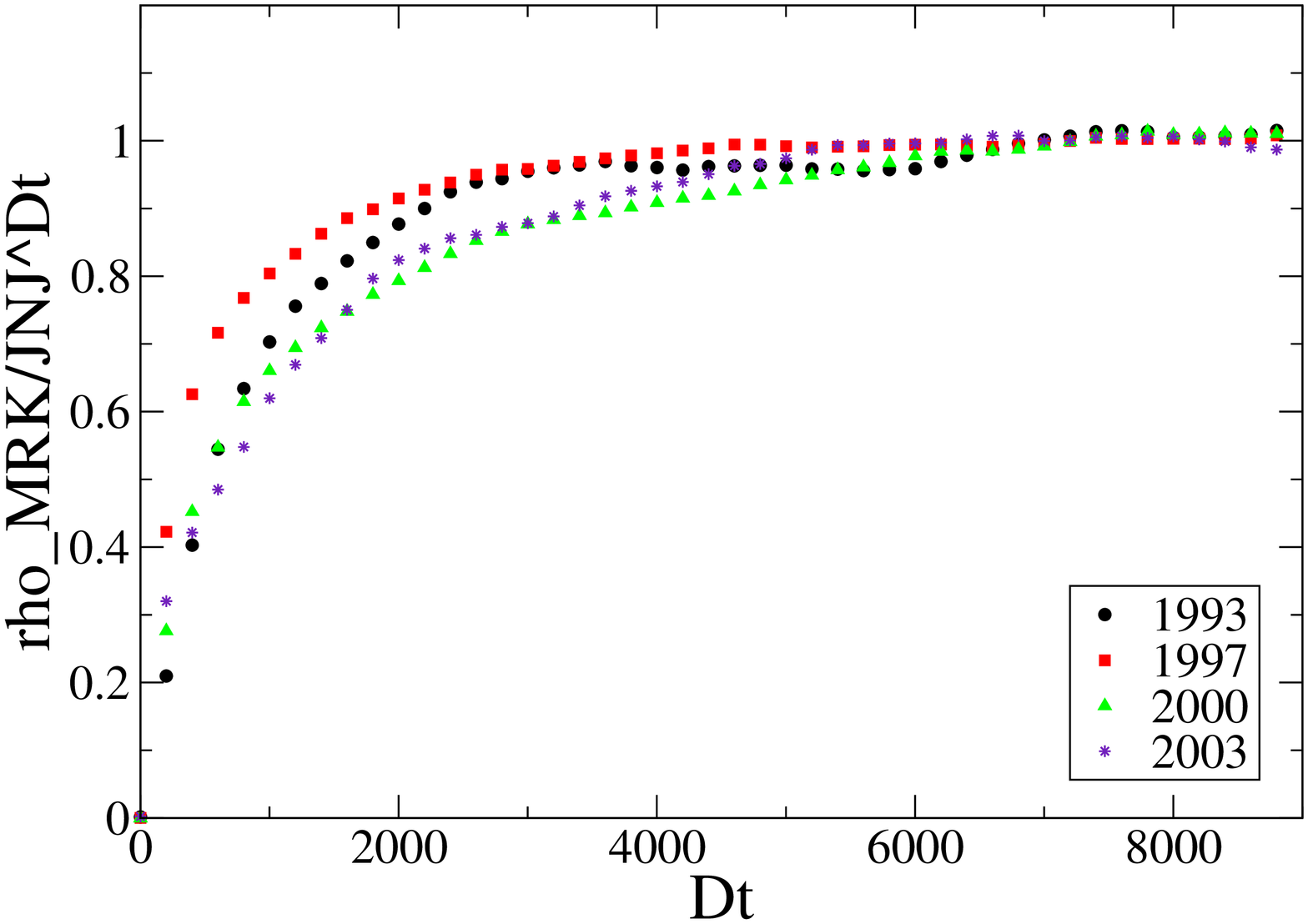}
\caption{The Epps curves scaled with the asymptotic cross-correlation values for the
  CAT/DE (top), KO/PEP (middle) and MRK/JNJ (bottom) pairs for the
  years 1993, 1997, 2000 and 2003. The scaled curves give a reasonable
  data collapse in spite of the considerably changing trading
  frequency, showing that the characteristic time of the Epps effect
  does not change with growing activity.}
\label{fig:epps_years_scaled}
\end{center}
\end{figure}

It is known that cross-correlation coefficients are not constant
through the years.  The asymptotic values of cross-correlations (long
sampling time scale) depend on the economical situation, the state of
the economic sectors that the pairs of stocks belong to, and several
other factors. We need to take this into account and try to extract
the effect of changing asymptotic cross-correlations from the Epps
phenomenon. In order to get comparable curves, we scaled the
cross-correlation curves with their asymptotic value: The latter was
defined as the mean of the cross-correlation coefficients for the
sampling time scales $\Delta t=6000$ seconds through $\Delta t=9000$
seconds, and the cross-correlations were divided by this value. Figure
\ref{fig:epps_years_scaled} shows the scaled curves for the same years and pairs
as Figure \ref{fig:epps_years}.

\begin{figure}[htb!]
\begin{center}
\psfrag{years}[t][b][4][0]{years}
\psfrag{ait}[b][t][4][0]{average intertrade time [sec]}
\includegraphics[angle=-90,width=0.50\textwidth]{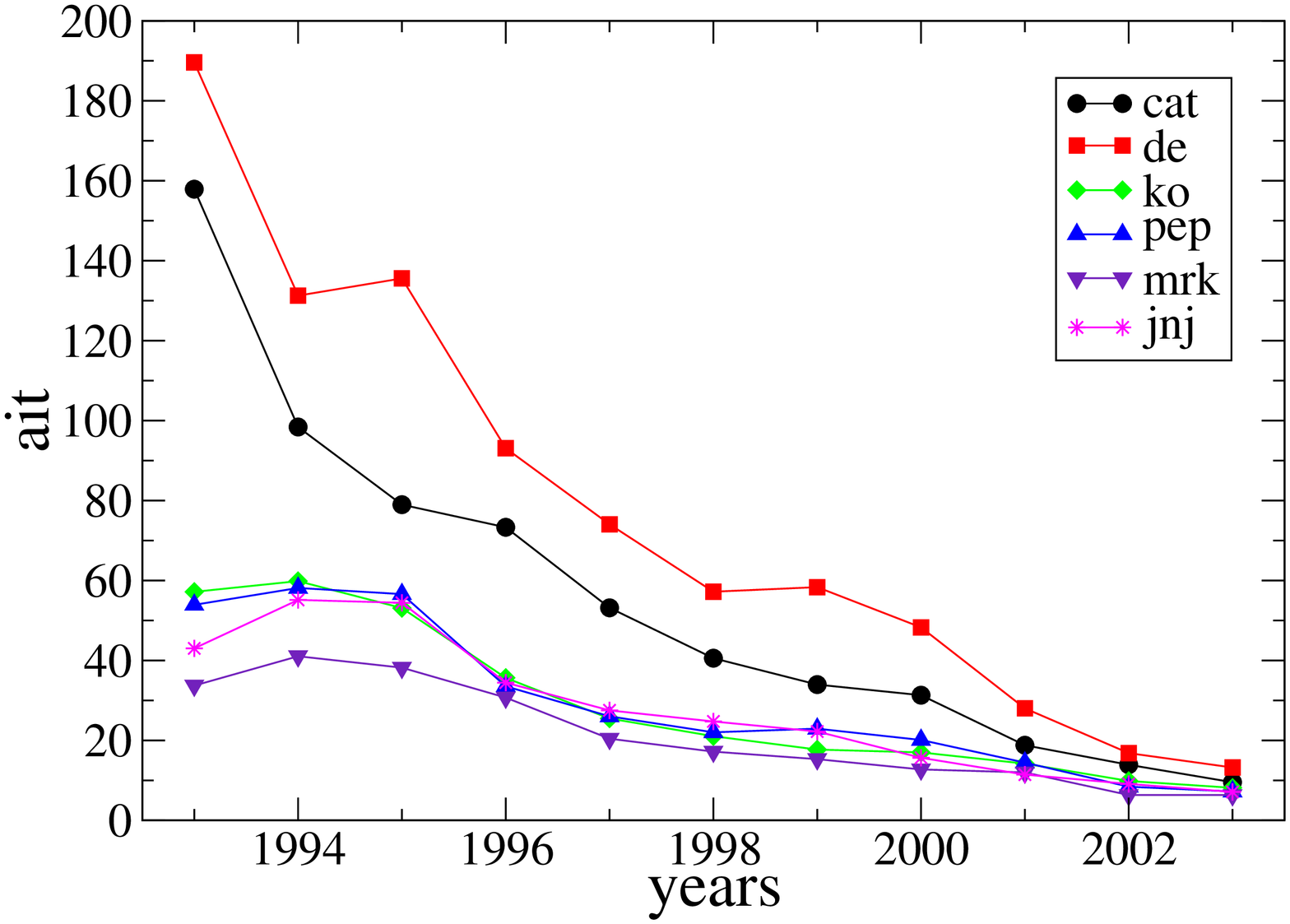}
\caption{The average intertrade time for the years 1993 to 2003 for some example 
stocks (CAT, DE, KO, PEP, MRK, JNJ). The activity was growing almost
monotonically.}
\label{fig:activity}
\end{center}
\end{figure}

The frequency of trades changed considerably in the last two decades: Trading
activity has grown almost monotonically, as it can be seen in Figure
\ref{fig:activity}.  This would infer the diminution of the Epps effect and a
much weaker decrease of the correlations as sampling frequency is
increased.  However, after scaling with the asymptotic
cross-correlation value, the curves give a reasonable data collapse
and no systematic trend can be seen. Surprisingly, as it can be seen
in Table \ref{table:t_E}, a rise of the trading frequency by a factor
of $\sim 5-10$ does not lead to a measurable reduction of the
characteristic time of the Epps effect (where we define the
characteristic time the time scale for which the cross-correlation
reaches the $1-e^{-1}$ rate of its asymptotic value).

These results show that explaining the Epps effect merely as a result
of the asynchronicity of ticks is not satisfactory. It is also
important to mention that not even the changing tick sizes for the
stocks (likely to change the arrival rate of price changes) alter the
characteristic time of the effect.

\begin{table}[htb!]

\caption{The characteristic time of the Epps effect for the years 1993, 1997, 2000 and 2003
measured in seconds for the stocks pairs: CAT/DE, KO/PEP and MRK/JNJ
(characteristic time was defined as the time scale for which the
cross-correlation value reaches the $1-e^{-1}$ rate of its asymptotic
value). No clear trends can be seen in the charactersitic time while
the activity is growing rapidly.}
\label{table:t_E}       

\begin{center}
\resizebox{0.50\textwidth}{!}{
\begin{tabular}{lllll}
    & \textbf{CAT/DE} & \textbf{KO/PEP} & \textbf{MRK/JNJ}  \\

\hline

\textbf{1993} \vline &   940   &   920   &   800 \\
\textbf{1997} \vline &   620   &   760   &   420 \\
\textbf{2000} \vline &  1320   &  1040   &   880 \\
\textbf{2003} \vline &   700   &   800   &  1060 \\
\noalign{\smallskip}\hline

\end{tabular}

}

\end{center}



\end{table}


\section{Decomposition of the cross-correlations}
\label{theory}
In this section we show calculations for the relation between the
value of cross-correlations on different time scales. We connect the
cross-correlation on a certain time scale ($\Delta t$) to lagged
autocorrelations and cross-correlations on smaller time scales
($\Delta t_0$).

Returns in a certain time window $\Delta t$ are mere sums of returns
in smaller, non-overlapping windows $\Delta t_{0}$, where $\Delta t$
is a multiple of $\Delta t_{0}$:

\begin{eqnarray}
\label{eq:ret_scale}
r_{\Delta t}(t)=\sum_{s=1}^{\Delta t / \Delta t_{0}}r_{\Delta
t_{0}}(t-\Delta t+s\Delta t_{0}).
\end{eqnarray}

Using this relationship the time average of the product of returns on
the large time scale ($\Delta t$) can be written in terms of the
averages on the short time scale ($\Delta t_0$) in the following way:

\begin{eqnarray}
\label{eq:ave_scale}
\left\langle r_{\Delta t}^{A}(t)r_{\Delta
    t}^{B}(t)\right\rangle=\frac{1}{T-\Delta t}\sum_{i=\Delta
    t}^{T}r_{\Delta t}^{A}(i)r_{\Delta t}^{B}(i)= \nonumber \\
    =\frac{1}{T-\Delta t}\sum_{i=\Delta t}^{T}\left(\sum_{s=1}^{\Delta
    t / \Delta t_{0}}r_{\Delta t_{0}}^{A}(i-\Delta t+s\Delta
    t_{0})\right)\left(\sum_{q=1}^{\Delta t / \Delta t_{0}}r_{\Delta
    t_{0}}^{B}(i-\Delta t+q\Delta t_{0})\right)= \nonumber \\
    =\sum_{s=1}^{\Delta t / \Delta t_{0}} \sum_{q=1}^{\Delta t /
    \Delta t_{0}} \left\langle r_{\Delta t_{0}}^{A}(i-\Delta t+s\Delta
    t_{0})r_{\Delta t_{0}}^{B}(i-\Delta t+q\Delta t_{0})\right\rangle.
\end{eqnarray}

We can see that on the right side of Equation \eqref{eq:ave_scale} the
lagged time average of return products appear on the short time scale,
$\Delta t_0$, i.e., the non-trivial part of the lagged
cross-correlations. Naturally in the case of $A=B$, we get the
relation for $\left\langle r_{\Delta t}(t)^{2}\right\rangle$.

In order to apply Equation \eqref{eq:ave_scale}, we need to have
information about the lagged autocorrelation and cross-correlation
functions. Writing out the sum in Eq.
\eqref{eq:ave_scale} we get:

\begin{eqnarray}
\label{eq:ave_assumption}
\left\langle r_{\Delta t}^{A}(t)r_{\Delta
    t}^{B}(t)\right\rangle=\sum_{x=-\frac{\Delta t}{\Delta
    t_0}+1}^{\frac{\Delta t}{\Delta t_0}-1}\left(\frac{\Delta
    t}{\Delta t_0}-|x|\right)\left\langle r_{\Delta
    t_0}^{A}(t)r_{\Delta t_0}^{B}(t+x\Delta t_0)\right\rangle,
\end{eqnarray}
and similarly
\begin{eqnarray}
\label{eq:ave_assumption_2}
\left\langle r_{\Delta t}^{A}(t)^2\right\rangle=\sum_{x=-\frac{\Delta t}{\Delta t_0}+1}^{\frac{\Delta t}{\Delta t_0}-1}\left(\frac{\Delta t}{\Delta
    t_0}-|x|\right)\left\langle r_{\Delta t_0}^{A}(t)r_{\Delta
    t_0}^{A}(t+x\Delta t_0)\right\rangle \nonumber \\
\left\langle r_{\Delta t}^{B}(t)^2\right\rangle=\sum_{x=-\frac{\Delta t}{\Delta t_0}+1}^{\frac{\Delta t}{\Delta t_0}-1}\left(\frac{\Delta t}{\Delta
    t_0}-|x|\right)\left\langle r_{\Delta t_0}^{B}(t)r_{\Delta
    t_0}^{B}(t+x\Delta t_0)\right\rangle.
\end{eqnarray}


Since the mean of returns is 1-2 orders of magnitude smaller than the
second moments in the correlation function, we can omit the
expressions $\left\langle r_{\Delta t}^{A}(t)\right\rangle
\left\langle r_{\Delta
t}^{B}(t+\tau)\right\rangle$ and $\left\langle r_{\Delta
t}(t)\right\rangle^{2}$ in Equation \eqref{eq:C}. As these terms are
of second order, this can even be done in case of slight price
trends. Hence Equation \eqref{eq:C} becomes:

\begin{eqnarray}
\label{eq:rho_assumption1}
\rho_{\Delta t}^{A/B}=\frac{\left\langle r_{\Delta t}^{A}(t)r_{\Delta
      t}^{B}(t)\right\rangle}{\sqrt{\left\langle r_{\Delta
        t}^{A}(t)^{2}\right\rangle\left\langle r_{\Delta
        t}^{B}(t)^{2}\right\rangle}}.
\end{eqnarray}

For simplicity, we introduce decay functions to describe lagged
correlations:

\begin{eqnarray}
\label{eq:def_decay}
f_{\Delta t_0}^{A/B}(x\Delta t_0)=\frac{\left\langle r_{\Delta
t_0}^{A}(t)r_{\Delta t_0}^{B}(t+x\Delta
t_0)\right\rangle}{\left\langle r_{\Delta t_0}^{A}(t)r_{\Delta
t_0}^{B}(t)\right\rangle},
\end{eqnarray}
defined for both positive and negative $x$ values, and similarly
$f_{\Delta t_0}^{A/A}(x\Delta t_0)$ and $f_{\Delta t_0}^{B/B}(x\Delta
t_0)$.  Thus the correlation can be written in the following form:

\begin{eqnarray}
\label{eq:data_formula1}
\rho_{\Delta t}^{A/B}=\Bigg(\sum_{x=-\frac{\Delta t}{\Delta t_0}+1}^{\frac{\Delta t}{\Delta t_0}-1}\left(\frac{\Delta t}{\Delta
      t_0}-|x|\right)f_{\Delta t_0}^{A/B}(x\Delta t_0)\left\langle
      r_{\Delta t_0}^{A}(t)r_{\Delta
      t_0}^{B}(t)\right\rangle\Bigg)\times \nonumber \\
\Bigg(\sum_{x=-\frac{\Delta t}{\Delta t_0}+1}^{\frac{\Delta t}{\Delta t_0}-1}\left(\frac{\Delta t}{\Delta
      t_0}-|x|\right)f_{\Delta t_0}^{A/A}(x\Delta t_0)\left\langle
      r_{\Delta t_0}^{A}(t)^2\right\rangle\Bigg)^{-1/2}\times
      \nonumber \\
\Bigg(\sum_{x=-\frac{\Delta t}{\Delta t_0}+1}^{\frac{\Delta t}{\Delta t_0}-1}\left(\frac{\Delta t}{\Delta
    t_0}-|x|\right)f_{\Delta t_0}^{B/B}(x\Delta t_0)\left\langle
    r_{\Delta t_0}^{B}(t)^2\right\rangle\Bigg)^{-1/2}.
\end{eqnarray}

Hence

\begin{eqnarray}
\label{eq:data_formula2}
\rho_{\Delta t}^{A/B}=\Bigg(\sum_{x=-\frac{\Delta t}{\Delta t_0}+1}^{\frac{\Delta t}
    {\Delta t_0}-1}\left(\frac{\Delta t}{\Delta
    t_0}-|x|\right)f_{\Delta t_0}^{A/B}(x\Delta t_0)\Bigg)\times
    \nonumber \\
\Bigg(\sum_{x=-\frac{\Delta t}{\Delta t_0}+1}^{\frac{\Delta t}{\Delta t_0}-1}\left(\frac{\Delta t}{\Delta
      t_0}-|x|\right)f_{\Delta t_0}^{A/A}(x\Delta
      t_0)\Bigg)^{-1/2}\times \nonumber \\
\Bigg(\sum_{x=-\frac{\Delta t}{\Delta t_0}+1}^{\frac{\Delta t}
    {\Delta t_0}-1}\left(\frac{\Delta t}{\Delta
    t_0}-|x|\right)f_{\Delta t_0}^{B/B}(x\Delta
    t_0)\Bigg)^{-1/2}\rho_{\Delta t_0}^{A/B}.
\end{eqnarray}

This way we obtained an expression of the cross-correlation
coefficient for any sampling time scale, $\Delta t$, by knowing the
coefficient on a shorter sampling time scale, $\Delta t_0$, and the
decay of lagged autocorrelations and cross-correlations on the same
shorter sampling time scale (given that $\Delta t$ is multiple of
$\Delta t_0$). Our method is to measure the correlations and fit their
decay functions on a certain short time scale and compute the Epps
curve using the above formula.

\section{Model calculations}
\label{model}

In this section we demonstrate the decomposition process on computer
generated time series which should mimic two correlated return
series. We will demonstrate the Epps effect and see how the
decomposition works for these controlled cases. Our aim is to show
that in case of generated ``price'' series, the decomposition process
leads to a very good description of the time scale dependence of the
cross-correlation coefficients. More discussion and details on the
analytic treatment of the model can be found in
Ref. \cite{firenze2007}.

\begin{figure}[htb!]
\begin{center}
\psfrag{time}[t][b][4][0]{time}
\psfrag{price}[b][t][4][0]{"price"}
\psfrag{W(t)}[][][2.5][0]{W(t)}
\psfrag{pA(t)}[][][2.5][0]{$p^{A}(t)$}
\psfrag{pB(t)}[][][2.5][0]{$p^{B}(t)$}
\includegraphics[angle=-90,width=0.50\textwidth]{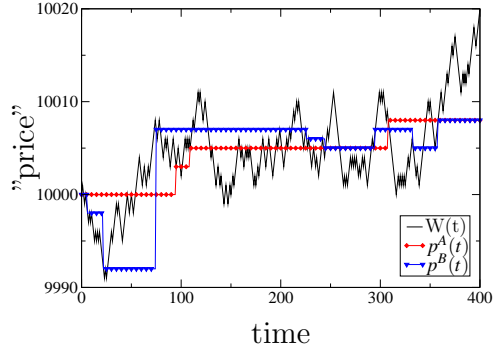}
\caption{A snapshot of the model with exponentially distributed waiting times. 
  The original random walk is shown with lines (black), the two
  sampled series with dots and lines (red) and triangles and lines
  (blue).}
\label{fig:model}
\end{center}
\end{figure}

To mimic some properties of financial data, we simulate two correlated
but asynchronous price time series. As a first step we generate a core
random walk with unit steps up or down in each second with equal
possibility ($W(t)$).  Second we sample the random walk, $W(t)$, twice
independently with waiting times drawn from an exponential
distribution.  This way we obtain two time series ($p^A(t)$ and
$p^B(t)$), which are correlated since they are sampled from the same
core random walk, but the steps in the two walks are asynchronous. The
core random walk is:

\begin{eqnarray}
\label{eq:model_def1}
W(t)= W(t-1)+\varepsilon(t), \nonumber \\
\end{eqnarray}
where $\varepsilon(t)$ is $\pm 1$ with equal probability (and $W(0)$
is set high in order to avoid negative values). The two price time
series are determined by

\begin{eqnarray}
\label{eq:model_def2}
p^A(t_i)=\Bigg\{\begin{array}{ll} 
W(t_i) & \textrm{if } t_i=\sum_{k=1}^{i}X_{k} \\
p(t_i-1) & \textrm{otherwise}
\end{array} \nonumber \\
p^B(t_i)=\Bigg\{\begin{array}{ll} 
W(t_i) & \textrm{if } t_i=\sum_{k=1}^{i}Y_{k} \\
p(t_i-1) & \textrm{otherwise}
\end{array}.
\end{eqnarray}
from the core random walk, where, $i=1,2,\cdots$, and $X_k$ and $Y_k$
are drawn from an exponential distribution:
\begin{eqnarray}
\label{eq:exp}
\mathbb{P}(y)=\Bigg\{\begin{array}{ll} 
\lambda e^{-\lambda y} & \textrm{if } y \ge 0 \\
0 & y<0
\end{array}
\end{eqnarray}
with parameter $\lambda=1/60$.  A snapshot as an example of the
generated time series with exponentially distributed waiting times can
be seen on Figure
\ref{fig:model}.

As a next step we create the logarithmic return time series
($r_{\Delta t}^A(t)$ and $r_{\Delta t}^B(t)$) of $p^A(t)$ and $p^B(t)$
as defined by Equation \eqref{eq:ret}. In case of the random walk
model of price changes we know that $\left\langle r_{\Delta
t}^{A}(t)\right\rangle=\left\langle r_{\Delta
t}^{B}(t)\right\rangle=0$ without having to make any assumptions.  Of
course having a random walk model, the autocorrelation function of the
steps is zero for all non-zero time lags:

\begin{eqnarray}
\label{eq:autodecay}
f_{\Delta t_0}^{A/A}(x\Delta t_0)=f_{\Delta t_0}^{B/B}(x\Delta
t_0)=\delta_{x,0},
\end{eqnarray}
thus 

\begin{eqnarray}
\label{eq:ave_assumption2}
\left\langle r_{\Delta t}^{A}(t)^2\right\rangle=\frac{\Delta t}{\Delta
  t_{0}}\left\langle r_{\Delta t_0}^{A}(t)^2\right\rangle \nonumber \\
\left\langle r_{\Delta t}^{B}(t)^2\right\rangle=\frac{\Delta t}{\Delta
  t_{0}}\left\langle r_{\Delta t_0}^{B}(t)^2\right\rangle.
\end{eqnarray}
Hence the cross-correlation can be written in the following form:

\begin{eqnarray}
\label{eq:model_formula}
\rho_{\Delta t}^{A/B}=\frac{\Delta t_0}{\Delta t}\Bigg(
\sum_{x=-\frac{\Delta t}{\Delta t_0}+1}^{\frac{\Delta t}{\Delta t_0}-1}\left(\frac{\Delta t}{\Delta
    t_0}-|x|\right)f_{\Delta t_0}^{A/B}(x\Delta t_0)\left\langle
    r_{\Delta t_0}^{A}(t)r_{\Delta
    t_0}^{B}(t)\right\rangle\Bigg)\times \nonumber \\
\Bigg(\left\langle r_{\Delta t_0}^{A}(t)^2\right\rangle\left\langle
    r_{\Delta t_0}^{B}(t)^2\right\rangle\Bigg)^{-1/2}= \nonumber\\
    =\frac{\Delta t_0}{\Delta t}\sum_{x=-\frac{\Delta t}{\Delta
    t_0}+1}^{\frac{\Delta t}{\Delta t_0}-1}\left[\left(\frac{\Delta
    t}{\Delta t_0}-|x|\right)f_{\Delta t_0}^{A/B}(x\Delta
    t_0)\right]\rho_{\Delta t_0}^{A/B}.
\end{eqnarray}

In the model case we set the smallest time scale $\Delta t_0 =1$ time
step.  It can be shown \cite{firenze2007} that in the case of
$\lambda\ll 1$ (small density of ticks) the exact analytical
expression for the cross-correlations is identical to
\eqref{eq:model_formula} with an exponential decay function:

\begin{eqnarray}
\label{eq:decay}
f_{\Delta t_0}^{A/B}(x\Delta t_0)=e^{-\lambda \Delta t_0 |x|},
\end{eqnarray}

where $\lambda$ is the parameter of the original exponential
distribution used for sampling. Further results and exact computations
of the cross-correlations for the model can be found in
Ref. \cite{firenze2007}.

\begin{figure}[htb!]  \begin{center} \psfrag{Dt}[t][b][4][0]{$\Delta
 t$ [simulation steps]} \psfrag{rho_Dt}[b][t][4][0]{$\rho_{\Delta
 t}^{A/B}$}
 \includegraphics[angle=-90,width=0.50\textwidth]{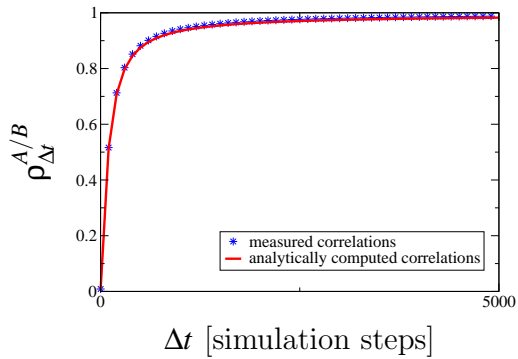}
 \caption{The measured and the computed cross-correlation coefficients
 using exponential decay function as a function of sampling time scale
 for the simulated time series with exponentially distributed waiting
 times. The analytic fit is in very good agreement with the Epps
 curve.}  \label{fig:model_epps_fit} \end{center} \end{figure}

Figure \ref{fig:model_epps_fit} shows the computed cross-correlations
of the generated time series on several sampling time scales and the
computed cross-correlations using Equation \eqref{eq:model_formula}
and the exponential decay function \eqref{eq:decay}. The two curves
are in very good agreement showing that the decomposition procedure is
able to well capture the Epps effect for generated time series.

\section{Application of the theory to the data}
\label{application}
In this section we discuss the properties of the decay functions in
case of real world data, and inserting them into Equation
\eqref{eq:data_formula2} we derive analytical fits for the measured
Epps curves.

\subsection{Decay functions}
\label{decay}

As discussed, we measure the equal-time cross-correlations and the
decay of cross and autocorrelations on a certain short sampling time
scale and from these we obtain the value of equal-time
cross-correlations on larger sampling time scales. To do this, in case
of the toy model, we had the possibility of using the smallest time
scale available in the generated data as $\Delta t_0$, i.e., the
resolution being one simulation step. When studying data from real
world markets, one has to make restrictions. As being the highest
resolution commonly used in financial analysis, it would be plausible
to choose windows of one second as $\Delta t_0$. However on this time
scale one is only able to measure noise, no valid correlations and
decay functions can be found. Thus we had to use less dense data for
the smallest sampling time scale: in the results shown below we set
$\Delta t_0=120$ seconds. Using this resolution we get an acceptable
signal-to-noise ratio and we hope not to lose too much information
compared to higher frequencies.

To avoid new parameters in the model we use the raw decay functions in
the formula \eqref{eq:data_formula2}, without fitting them. Since it
is an empirical approach to determine the decay functions for real
data, we have to distinguish the signal from the noise in the decay
functions.  Concerning the sensitivity from the input (decay function)
we observed that the results are quite robust against little changes
in the input functions, however the noise in the tail can cause
significant deviations. According to this we take into account the
decay functions for correlations only for short time lags. For the
decay of the cross-correlations we take into account the function only
up to the time lag where the decaying signal reaches zero for the
first time, for larger lags we assume it to be zero.
For the decay of autocorrelations we take into the account the
function only up to the time lag where after the negative overshoot of
the beginning it decays to zero from below for the first time, for
larger lags we assume it to be zero.

\begin{figure}[htb!]
\begin{center}
\psfrag{tau}[t][b][4][0]{$x\Delta t_0$ [sec]}
\psfrag{f^KO/KO_Dtnull(tau)}[b][t][4][0]{$f^{KO/KO}_{\Delta t_{0}}(x\Delta t_0)$}
\psfrag{f^PEP/PEP_Dtnull(tau)}[b][t][4][0]{$f^{PEP/PEP}_{\Delta t_{0}}(x\Delta t_0)$}
\psfrag{f^KO/PEP_Dtnull(tau)}[b][t][4][0]{$f^{KO/PEP}_{\Delta t_{0}}(x\Delta t_0)$}
\includegraphics[angle=-90,width=0.50\textwidth]{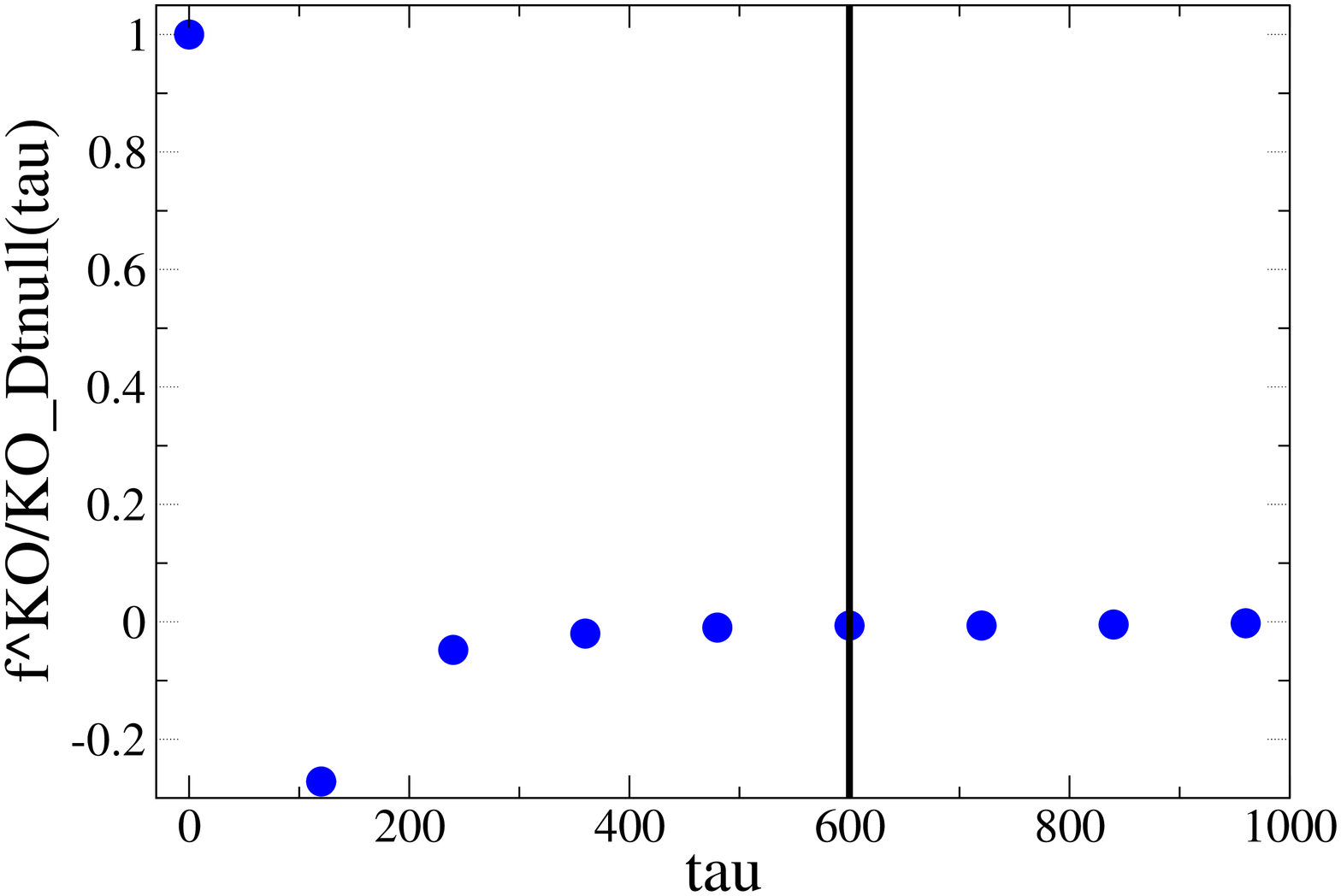}  
\includegraphics[angle=-90,width=0.50\textwidth]{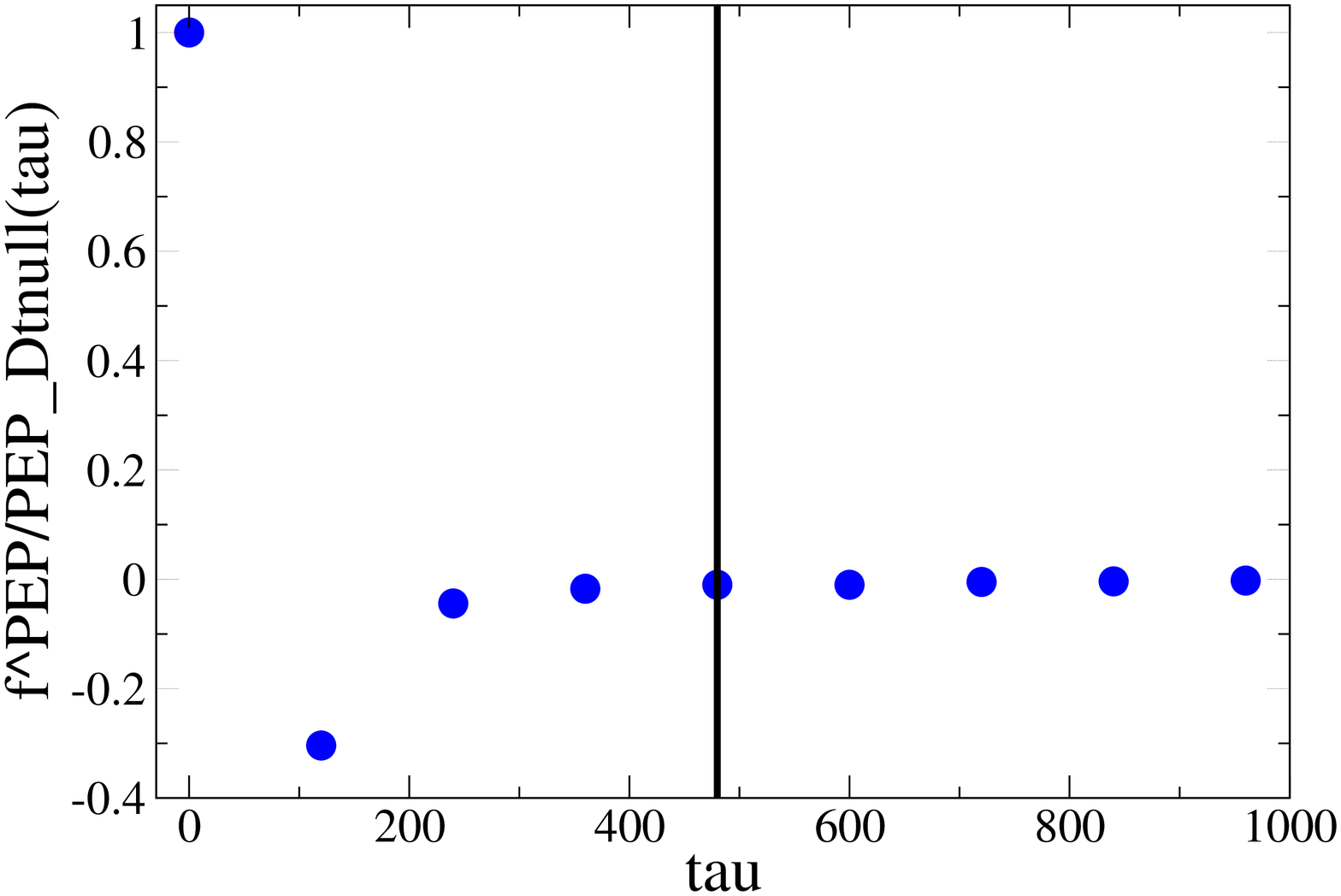} 
\includegraphics[angle=-90,width=0.50\textwidth]{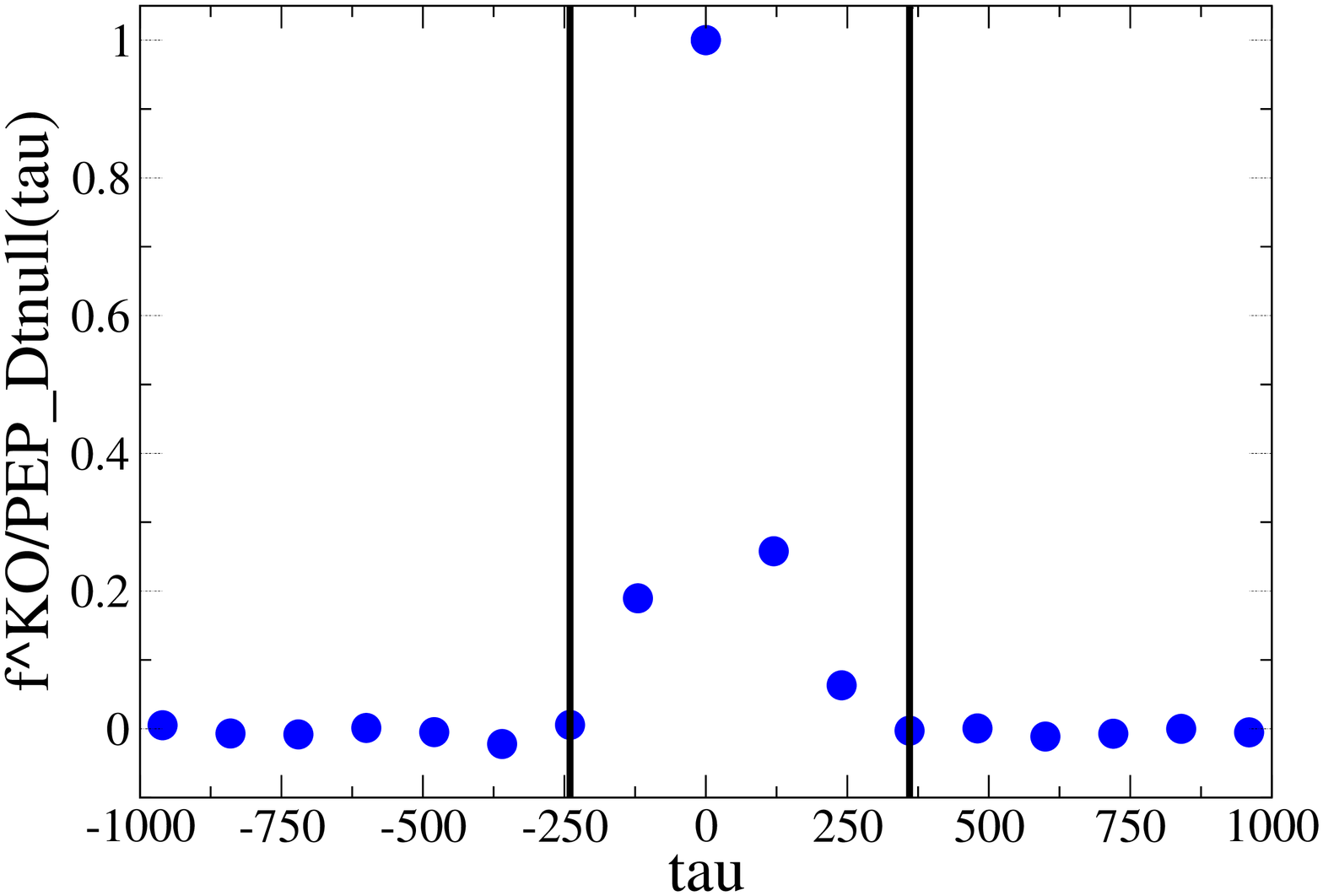}
\caption{Top: The decay of lagged autocorrelations for KO. Middle: The decay
  of lagged autocorrelations for PEP. Bottom: The decay of lagged
  cross-correlations for KO/PEP pair. A vertical lines show the
  threshold up to which we take the decays into account, for larger
  lags we assume them to be zero. Sampling time scale is $\Delta
  t_0=120$ seconds on all three plots.}
\label{fig:decay}
\end{center}
\end{figure}

Figure \ref{fig:decay} shows an example of the decay functions in case
of the stock pair KO/PEP (for other pairs the decay functions are very
much similar).  The plot shows the decay functions up to the time lags
of 1000 seconds, with a vertical line showing how long we take the
empirical decays into account.  We can see that the time lag for which
the decay functions disappear is in the order of a few minutes. In
fact in case of all stock pairs studied we found the decay
disappearing after 5--15 minutes.

\subsection{Fits}
\label{fit}
\begin{figure}[htb!]
\begin{center}
\psfrag{Dt}[t][b][4][0]{$\Delta t$ [sec]}
\psfrag{rho^CAT/DE_Dt}[b][t][4][0]{$\rho^{CAT/DE}_{\Delta t}$}
\psfrag{rho^KO/PEP_Dt}[b][t][4][0]{$\rho^{KO/PEP}_{\Delta t}$}
\hbox{
\includegraphics[angle=-90,width=0.50\textwidth]{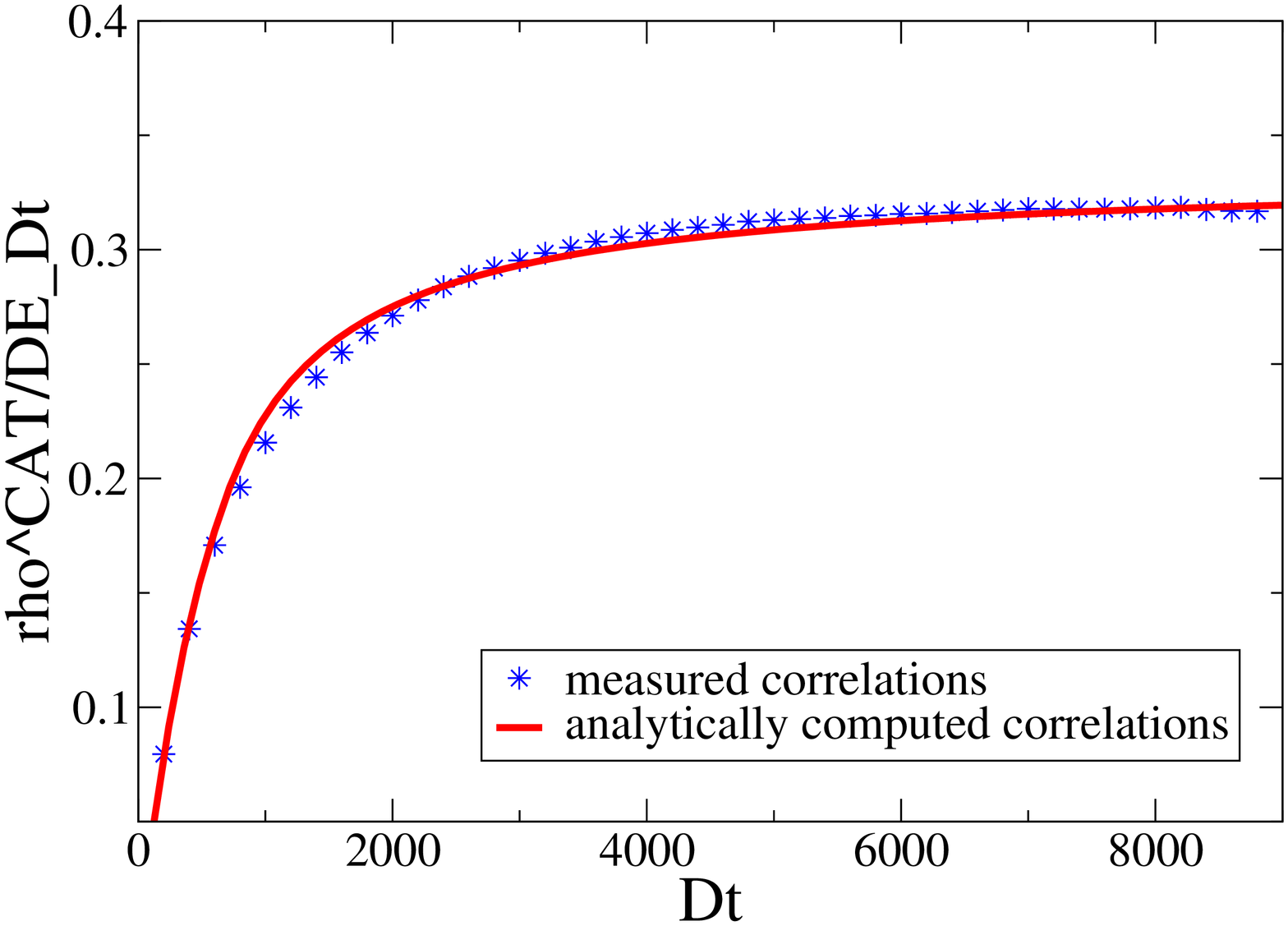}  
\includegraphics[angle=-90,width=0.50\textwidth]{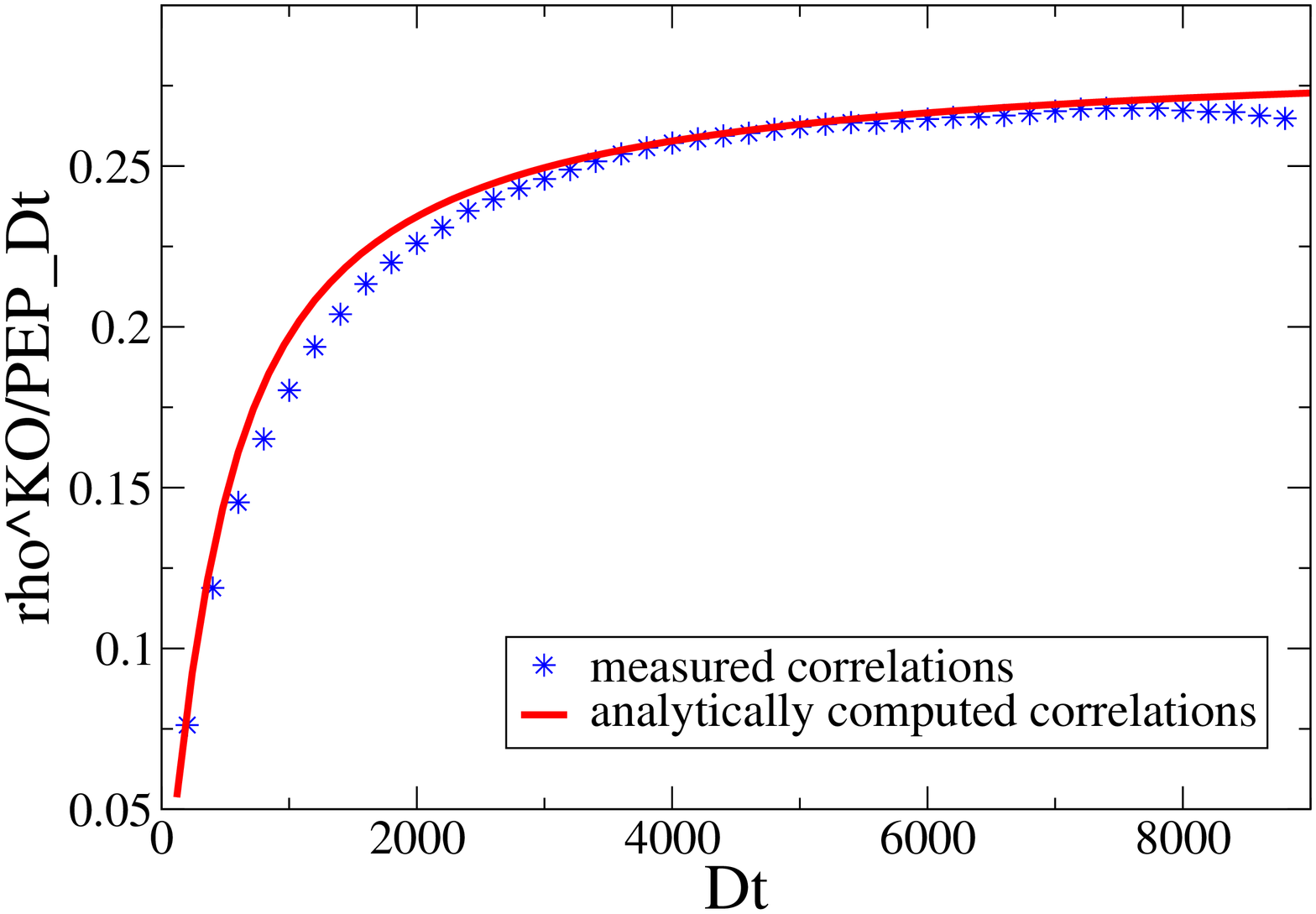} 
}
\caption{The measured and the analytically computed cross-correlation coefficients
  as a function of sampling time scale for the pairs CAT/DE and
  KO/PEP. Note that using only the autocorrelations and
  cross-correlations measured on the smallest time scale ($\Delta
  t_0=120$ seconds) we are able to give reasonable fits to the
  cross-correlations on all time scales. Details on the goodness
  parameter of the measured and computed correlations can be found in
  Table \ref{table:goodness}.}
\label{fig:fit1}
\end{center}
\end{figure}

\begin{figure}[htb!]
\begin{center}
\psfrag{Dt}[t][b][4][0]{$\Delta t$ [sec]}
\psfrag{rho^S/WMT_Dt}[b][t][4][0]{$\rho^{WMT/S}_{\Delta t}$}
\psfrag{rho^GE/MOT_Dt}[b][t][4][0]{$\rho^{GE/MOT}_{\Delta t}$}
\hbox{
\includegraphics[angle=-90,width=0.50\textwidth]{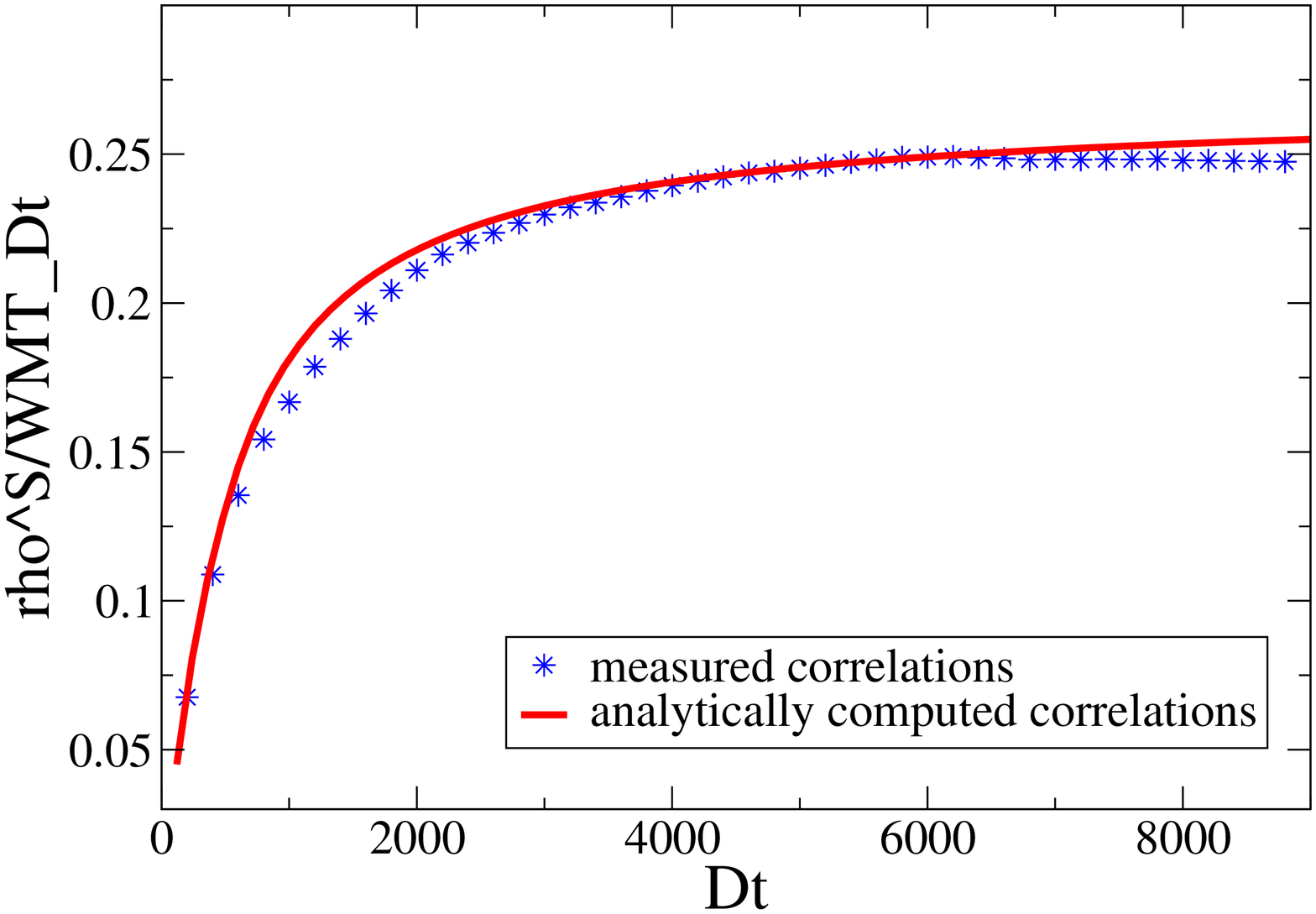}  
\includegraphics[angle=-90,width=0.50\textwidth]{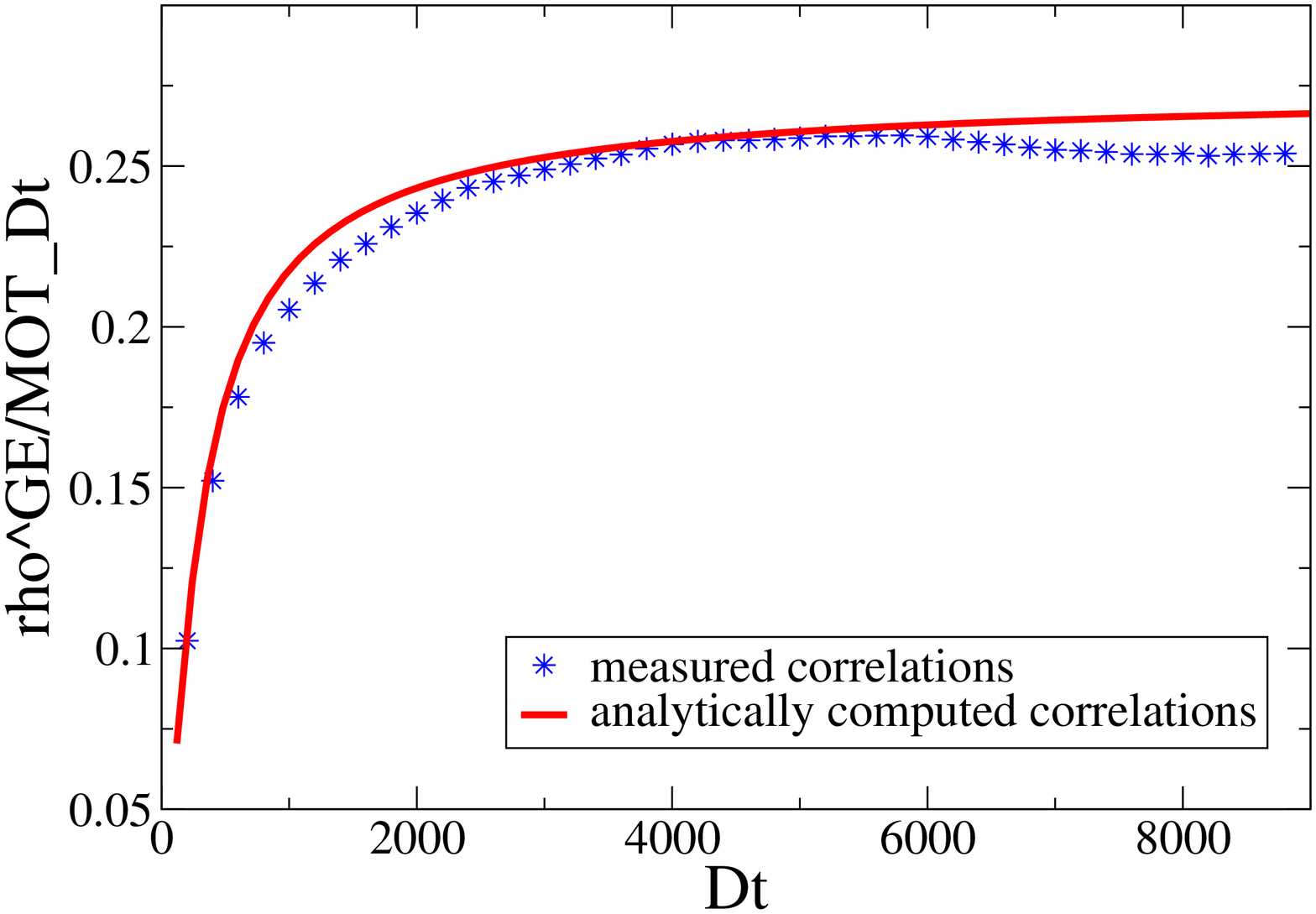} 
}
\caption{The measured and the analytically computed cross-correlation coefficients
  as a function of sampling time scale for the pairs WMT/S and
  GE/MOT. Details on the goodness parameter of the measured and
  computed correlations can be found in Table \ref{table:goodness}.}
\label{fig:fit2}
\end{center}
\end{figure}

\begin{figure}[htb!]
\begin{center}
\psfrag{Dt}[t][b][4][0]{$\Delta t$ [sec]}
\psfrag{rho^MRK/JNJ_Dt}[b][t][4][0]{$\rho^{MRK/JNJ}_{\Delta t}$}
\psfrag{rho^GE/IBM_Dt}[b][t][4][0]{$\rho^{GE/IBM}_{\Delta t}$}
\hbox{
\includegraphics[angle=-90,width=0.50\textwidth]{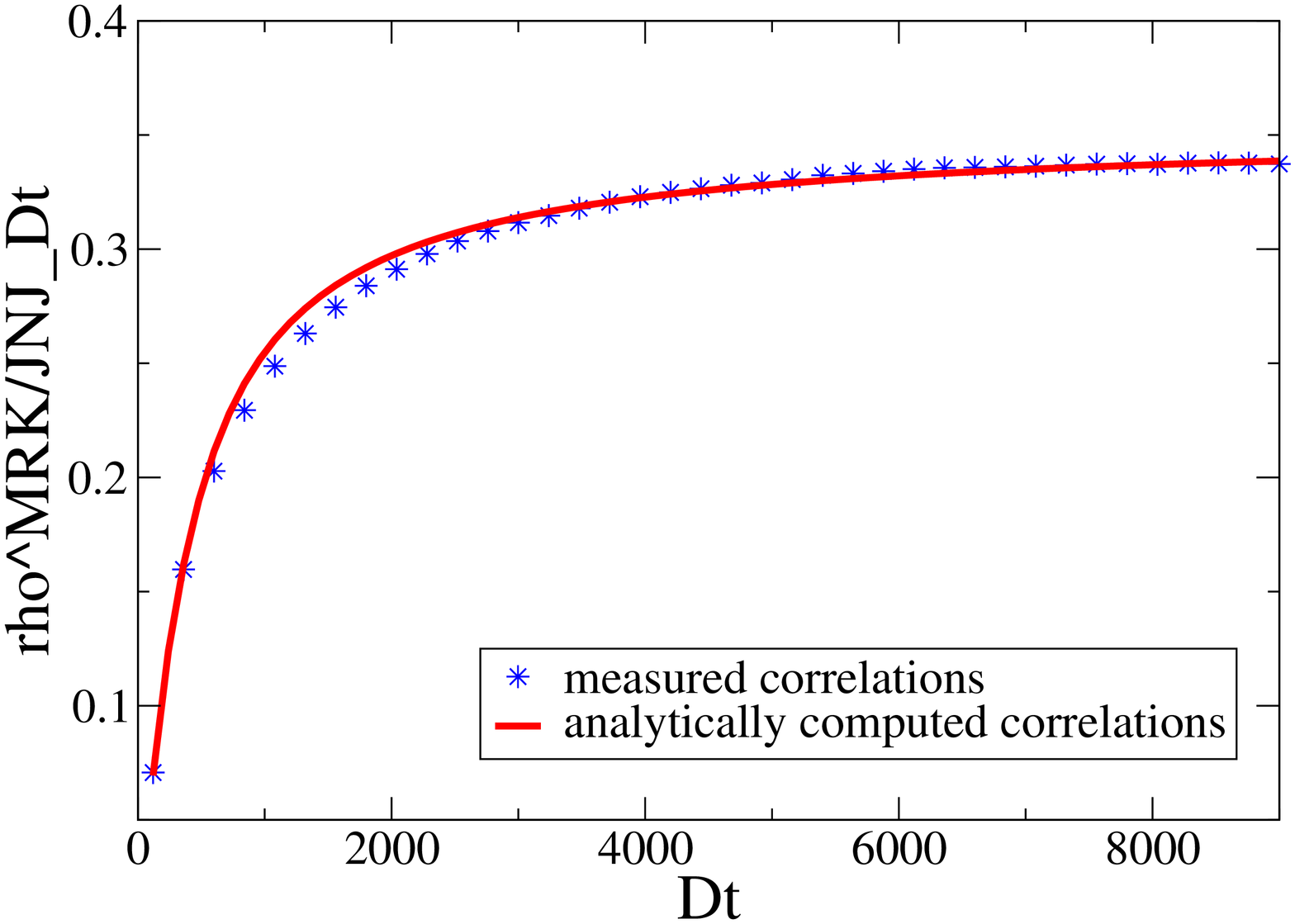}
\includegraphics[angle=-90,width=0.50\textwidth]{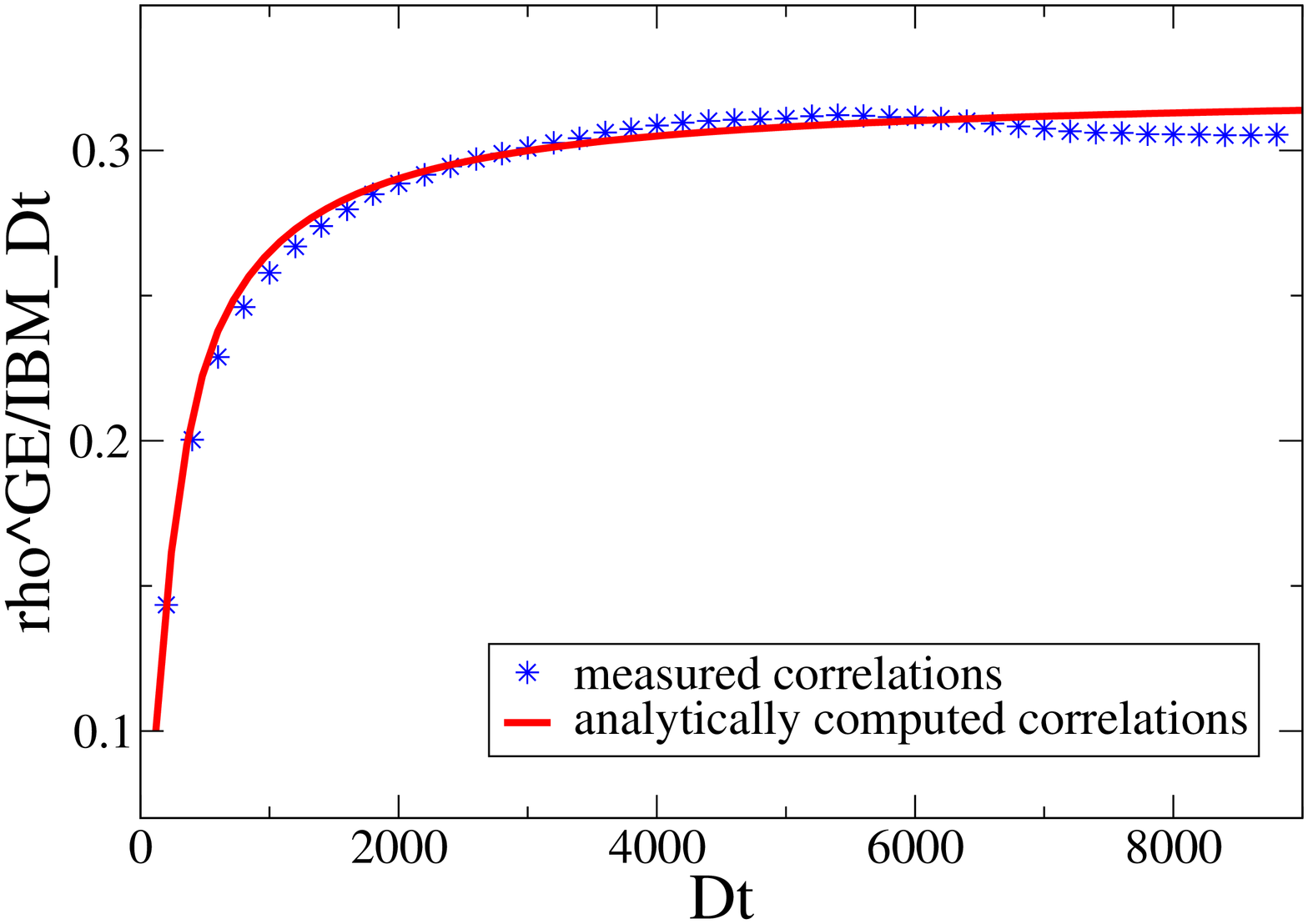}
}
\caption{The measured and the analytically computed cross-correlation coefficients
  as a function of sampling time scale for the pairs MRK/JNJ and GE
  /IBM. Details on the goodness parameter of the measured and computed
  correlations can be found in Table \ref{table:goodness}.}
\label{fig:fit3}
\end{center}
\end{figure}

Inserting the empirical decay of lagged autocorrelations and
cross-correlations on the short time scale into the formula of
Equation \eqref{eq:data_formula2}, we compare the computed and the
measured Epps curves. Figures \ref{fig:fit1}, \ref{fig:fit2}
and \ref{fig:fit3} show these plots for a few example stock pairs.

One can see, that the fits are able to describe the change of
cross-correlation with increasing sampling time scale. Note, that as
it has been shown in Section
\ref{theory}, in the analytical formula only the autocorrelations and cross-correlations on the smallest
time scale ($\Delta t_0$) and the decay functions are taken into
account as input to compute the cross-correlations on all other time
scales, no additional parameters are used.

To show a broader set of results, we introduce a goodness parameter
for the agreement between the measured and the analytically determined
Epps curves. We define the goodness parameter as the absolute error
between the measured and the analytically computed points:

\begin{eqnarray}
g(\Delta t)=100\frac{|\rho^{measured}_{\Delta t}-\rho^{computed}_{\Delta t}|}{\rho^{measured}_{\Delta t}}.
\end{eqnarray}

Table \ref{table:goodness} shows the maximum, the mean and the median
of the goodness parameters for a broader set of stocks. The results
show that the absolute mean error is very low, with a maximum around 7
percents and both a mean and a median around 2 percents. It is
important to mention, that the maximal error is usually found for high
frequency scales, for longer time scales and especially for the
asymptotic correlation value the aggrement is very good.
 
\begin{table}[htb!]

\caption{The maximum, the mean and the median of the goodness parameters a 
broader set of stocks. The results show that the absolute mean error
is low. Note that the maximal absolute error in general occurs for
high frequency scales.}
\label{table:goodness}       

\begin{center}
\resizebox{0.6\textwidth}{!}{
\begin{tabular}{lllll}
\textbf{stock pair} & \textbf{max} [\%] & \textbf{mean} [\%] & \textbf{median} [\%] \\

\hline

CAT/DE & 4.81 & 1.26  & 0.94   \\
KO/PEP & 10.67 & 2.46  & 1.23   \\
WMT/S & 7.66 & 2.32  & 1.74   \\
GE/MOT & 6.43 & 3.29  & 3.44   \\
MRK/JNJ & 5.26 & 1.51  & 0.95   \\
GE/IBM & 3.90 & 1.57  & 1.12   \\
PG/CL & 6.05 & 1.81  & 1.24   \\
MRK/PFE & 4.76 & 1.42  & 1.32   \\
AVP/CL & 10.37 & 7.75  & 9.53   \\
DD/DOW & 8.84 & 2.05  & 1.49   \\
DD/MMM & 5.76 & 2.17  & 1.93   \\
MOT/VOD & 9.73 & 2.57  & 1.82   \\
DIS/GE & 5.78 & 1.54  & 0.97   \\

\hline

average & 6.92 & 2.4  &  2.1   \\

\noalign{\smallskip}\hline

\end{tabular}

}

\end{center}
\end{table}

These results show that the growing cross-correlations with decreasing
sampling frequency are due to finite time decay of the lagged
autocorrelations and cross-correlations in the high frequency sampled
data.

The finite decay of the cross-correlations on the short time scale
($\Delta t_0$) is not caused by difference in the capitalisation of
the two stocks or functional dependencies between them. Instead, it is
an artifact of the market microstructure. Reaction to a certain piece
of news is usually spread out on an interval of a few minutes for the
stocks
\cite{Dann1977,PatellWolfson1984,JenningsStarks1985,BarclayLitzenberger1988,Kim1997,Busse2002,Chordia2005,Chordia2008,dacorogna_book,almeida1998}
due to human trading nature, thus not scaling with activity, with
ticks being distributed more or less randomly.  This means that
correlated returns are spread out for this interval (asynchronously),
causing non zero lagged cross-correlations on the short time scale and
thus the Epps effect.  This way, as stated by Ref.  \cite{reno2003},
the asynchronicity is indeed important in describing the Epps effect
but only in promoting the lagged correlations.  (Even in case of
completely synchronous, but randomly spread ticks we could have the
finite decay of lagged correlations on short time scale, and hence the
Epps effect.)

\section{Discussion}
\label{discussion}

In our study we examined the causes of the Epps effect, the dependence
of stock return cross-correlations on sampling time scale.  We showed
that explaining the effect solely through asynchronicity of price
ticks is not satisfactory.  When scaling the Epps curves with their
asymptotic value for different years, we get a reasonable data
collapse and a growing activity of the order $\sim 5-10$ does not
affect the characteristic time of the Epps effect.

The main point of our calculations is that we connected the
cross-correlations on longer time scales to the lagged
autocorrelations and cross-correlations on any shorter time scale. We
demonstrated the idea of these calculations on a random walk
asynchronous model of prices, getting a very good agreement with the
cross-correlation curves.

Assuming the time average of stock returns to be zero we were able to
decompose the expression for the cross-correlation coefficient
deriving an analytical formula of the cross-correlations on any time
scale, given the decay of the autocorrelations and cross-correlations
on a certain short time scale.  With this analytical formula we were
able to give fits to the Epps curves of real stock pairs getting
acceptable results. The fits show that the Epps effect is caused by
the finite time decay of the lagged correlations in the high frequency
sampled data. The reason for the characteristic time not changing with
growing activity is a human time scale present in the phenomenon,
which does not scale with the changing inter-tick time. The finite
decay of lagged correlations on the short time scale is due to market
microstructure properties: different actors on the market have
different time horizons of interest resulting in the reactions to
certain pieces of news being spread out for a time interval of a few
minutes. The correlated returns ranging over this interval cause the
finite time decay of lagged correlations on the short time scale
resulting in the Epps effect. Our results do not contradict to the
earlier observations on the importance of asynchronicity in the
Epps-effect, however, its role has been put into a new perspective.

\section*{Acknowledgments}
Support by OTKA T049238 is acknowledged.

\end{document}